\documentclass[prl,aps]{revtex4}
\begin{document}

\title{Negative Pressures and Energies in Magnetized and  Casimir Vacua}

\author{H. P\'erez Rojas and E. Rodr\'iguez Querts}

\affiliation{Grupo de Fisica Teorica, Instituto de Cibern\'etica, Matem\'atica y F\'{\i}sica,\\
Calle E  No. 309, esq. a 15 Vedado, C. Habana,Cuba
\\ hugo@icmf.inf.cu, elizabeth@icmf.inf.cu}

\begin{abstract}We study the
electron-positron vacuum in a strong
magnetic
field $B$ in parallel with Casimir effect. Use is made of the energy-momentum tensor,
taken as the zero temperature and
zero density limit of the relativistic quantum statistical tensor. In both magnetic field and Casimir
cases
it is shown the arising of anisotropic pressures. In the
first case  the pressure
transversal to the field $B$ is negative, whereas along $B$  an usual
positive pressure arises. Similarly, in addition to the usual
negative Casimir pressure perpendicular to the plates, the
existence of a positive pressure along the plates is predicted.
The anisotropic pressures suggests a flow of the virtual particles in both cases.
By assuming regions of the
universe having random orientation of the lines of force,
cosmological consequences are discussed in the magnetic field case.

\end{abstract}
 \maketitle
\section{Introduction}

There
is some similarity among the effects of an
external
field and certain boundary conditions. For instance, this is the case of
virtual electron-positron pairs of vacuum under the action of an
external constant magnetic field, and the virtual photon field
modes inside the cavity formed by two parallel metallic plates
(Casimir effect \cite{Casimir},\cite{Milonni}. For more recent
developments
see \cite{Bordag}). Both problems bear
some interesting analogy,  the metallic plates keeping bounded the
motion of virtual photons perpendicular to the plates whereas the
external field bound electrons and positrons to Landau quantum
states in the plane perpendicular to it.

In both cases, by starting from the expression for the zero point energy
density
of vacuum, and after subtracting some divergent terms, a finite negative
energy
density is left. From it, one can obtain the corresponding energy-momentum
tensor
density (in what follows, when referring to the energy-momentum
tensor, we would mean its density). Its spatial components contain
both negative and positive pressures: in the magnetic field case,
an axially symmetric transverse negative pressure, and a positive
pressure along the magnetic field. In the Casimir case, in
addition to the well known negative pressure, it is found an
axially symmetric positive pressure inside the cavity and along
the plates.

The comparative study of these two QED vacua (in an independent
way, but following a common method) is interesting in itself,
exhibiting the property of anisotropic pressures, having negative
values in some directions, as well as negative energy density.

We assume the magnetic field characterized by the microscopic
magnetic field pseudo-vector $B_{i}=\epsilon_{ijk}{\cal F}_{jk}$
(where ${\cal F}_{jk}$ is the spatial part of the electromagnetic
field tensor ${\cal F}_{\mu \nu}$), leading to the breaking of the
spatial symmetry: the spinor wavefunctions and spectrum of charged
particles having axial symmetry, their motion being bounded
perpendicularly to $B$ \cite{Johnson}. Its characteristic length
is $\lambda_{L}(B)=\sqrt{\hbar c/2eB}$, and this is valid also for
the zero point modes of vacuum. In the case of the the well-known
Casimir effect it is produced when two parallel metallic plates
are placed in vacuum,  leading to the vanishing of the electric
field component tangential to the plates. The plates having
characteristic diameter  $a$ and separated by a distance $d$, with
$a>>d$ (we shall assume in what follows $a \to \infty$). Only
modes whose wave vector components perpendicular to the cavity are
integer multiples of $k_{03}=\pi/d$, are allowed inside it. This
makes the zero point electromagnetic modes inside the box axially
symmetric in momentum space.  In both problems there is a quantity
characterizing the symmetry breaking, (and the extension of the
wave functions in some direction). These quantities are
respectively, the pseudo-vector $B_i$, determining
$\lambda_{L}(B)$, and the basic vector momentum ${\cal P}_i={\cal
P}\delta_{3i}$, perpendicular to the plates (which are taken
parallel to the $x_1, x_2$ plane). Here ${\cal P}= \hbar k_{03}$
and the length $d$ characterizes the extension of the wave
function perpendicular to the plates.

In the magnetic field case \cite{Johnson}, the solution of the
Dirac equation for an electron (or positron) in presence of an
external magnetic field $B_j$ for, say, $j=3$, leads to the energy
eigenvalues,
\begin{equation}\label{eigenv}
\varepsilon_{n}=\sqrt{c^2 p_{3}^{2}+m^{2}c^4+2e\hbar cB
n},\label{smag}
\end{equation}
\noindent  where $n=0,1,2...$ are the Landau quantum numbers,
$p_{3}$ is the momentum component along the magnetic field ${\bf B
}$ and $m$  is the electron mass. The breaking of the spatial
symmetry due to the magnetic field is manifested in the spectrum
as an harmonic oscillator-like quantization of the energy in the
direction perpendicular to the field. It has the form $c^2
p_{\perp}^2=e\hbar c B(2n+1)$. This term combines with the spin
contribution $\pm e\hbar cB$, leading to the last term inside the
square root in (\ref{smag}). Below we will integrate over $p_3$
and sum over $n$ and extract the finite  contribution to the
vacuum energy. The system is degenerate with regard to the
coordinates of the orbit´s center \cite{Johnson}.

In the Casimir effect the motion of virtual photons perpendicular
to the plates is bounded and we have the photon energy eigenvalues,
\begin{equation}
\varepsilon_s=c\sqrt{p_1^2+p_2^2+({\cal P}s)^2}. \label{Cas}
\end{equation}
\noindent Due to the breaking of the rotational symmetry in
momentum space inside the cavity, only vacuum modes of discrete
momentum $p_3={\cal P}s$ where $s=0,\pm 1, \pm 2,...$ are allowed.
After taking the sum of all these modes and subtracting the
divergent part from it \cite{Milonni}, one gets a finite negative
term dependent on ${\cal P}$, which is the vacuum energy. From
this, it was shown by Casimir \cite{Casimir},\cite{Milonni} that a
negative pressure appears in between the plates and perpendicular
to them.

Starting from the analogy between (\ref{Cas}) and (\ref{smag}),
which have as a common property a discrete quantization of some of
its momentum components, it would be interesting to investigate in
parallel both physical systems, especially since a negative
pressure arises as due to the zero point
electron-positron vacuum energy in an external constant magnetic
field.

\section{Vacuum zero point energies}

\noindent The electron-positron zero point vacuum energy in an
external electromagnetic field was obtained by  Heisenberg and
Euler \cite{Euler}.  For the case of a pure magnetic field, the
calculation of the zero point energy in the tree level
approximation (the virtual particles interact with the external
field, but not among themselves) can be made by starting from the
spectrum (\ref{smag}). By summing over all degrees of freedom, the
resulting term would contain the contribution from the virtual
electron-positron pairs created and annihilated spontaneously in
vacuum and interacting with the field $B$. By subtracting a fourth
order divergent term, independent of $B$,
 and a logarithmically divergent
term, proportional to $B^2$, where both divergences are due  to
the short wavelength modes, one extracts finally the finite term,
\begin{equation} \Omega_{0e} =
\frac{\alpha B^2}{8\pi^2}\int_0^{\infty}e^{-B_c
x/B}\left[\frac{coth x}{x}
-\frac{1}{x^2}-\frac{1}{3}\right]\frac{d x}{x}, \label{EH}
\end{equation}
\noindent where $\alpha=e^2/\hbar c$ is the fine-structure
constant and $B_c=m^2 c^3/e\hbar=4.41 \cdot 10^{13}$G is the QED
critical magnetic field. (We observe that the addition of a
negative infinite term proportional to $B^2$ absorbs the classical
energy term $B^2/8\pi$). As the quantity in squared brackets in
(\ref{EH}) is negative, we have $\Omega_{0e}<0$.
(We want to stress at this point that our aim is to study both Casimir and
magnetic field problems as independent. Both problems were investigated
together to find the magnetic permeability in Casimir vacuum in
\cite{Cougo})

The density of energy $\Omega_{0C}$ for the Casimir problem may be
obtained directly from (\ref{Cas}) by summing over $s$. This is
mathematically equivalent to find the thermodynamical potential of
radiation at a temperature $T_{Cas} =\hbar c/2d$, according to
temperature quantum field theory methods  \cite{Fradkin}. One gets
a finite term, dependent on $d$, and a four dimensional divergence
independent of $d$. The finite energy density is,
\begin{equation}
\Omega_{0C}= -\frac{\pi^2 \hbar c}{720 d^4}=-\frac{c{\cal
P}^{4}}{720 \pi^2\hbar^3 }. \label{Milon}
\end{equation}
\noindent Returning to (\ref{EH}), as fields currently achieved in
laboratories are very small if compared with the critical field
$B_{c}$, in the limit $B<< B_c$ one can write,
\begin{equation}\label{magnetization2}
{\Omega_{0e}}\approx-\frac{\alpha
B^{4}}{360\pi^{2}B_c^2}=-\frac{\pi^{2} \hbar c}{5760 b^{4}},\label{h}
\end{equation}

where the characteristic parameter is $b(B)= \pi
\lambda_{L}^{2}/\lambda_{C}$. Here
  $\lambda_{L}$ is the magnetic wavelength defined previously
and $\lambda_{C}$ is the Compton wavelength
$\lambda_{C}=\hbar/mc$.  The energy density is then a
  function of
  the field dependent parameter $b(B)$.
The expression for (\ref{magnetization2}) looks similar to
(\ref{Milon})

Both $\Omega_{0e}$ and $\Omega_{0C}$ are relativistic negative
energies, which suggests that these vacua bears  negative masses,
which would lead
to repulsive gravity with ordinary matter. From these energy densities
we will obtain respectively
 the energy-momentum tensor expression of vacuum in a constant
 magnetic field, (as the limit of the corresponding expression for matter
given i.e. in
\cite{Chaichian}) and in Casimir effect, and in consequence, the magnetic
and Casimir pressures.

\section{The energy-momentum tensor}
For the sake of completeness and correspondence with the quantum
gases of charged \cite{Chaichian}, and neutral particles bearing a
magnetic moment \cite{Aurora}, we discuss in what follows the
vacuum energy-momentum tensor by starting from the quantum
relativistic matter tensor, which contains the contribution of
vacuum. Usually the calculations are made in Euclidean variables, where $x_4$ is taken as some "imaginary time", but vectors and
tensors will be written here by using covariant and contravariant indices.
We will consider the usual QED Lagrangian density $L$ at finite temperature
$T$ and with conserved number of fermions $N=Tr\gamma_0\int d^3 x \bar
\psi({\bf x})\psi({\bf x})$, having associated chemical potential
$\mu$. One can write in an arbitrary moving frame
the density matrix as
\begin{equation}
\rho= e^{-\beta( u_\nu {\cal P}^\nu-\mu u_\nu J^\nu) }
\end{equation}
where ${\cal P}^\nu$ is the momentum four-vector,  $J^\nu=N
u^\nu$, and $u_\nu$ is the four-velocity of the medium.
From $\rho$, in the rest system, where ${\cal L}=\int d^3 x L$, one gets the effective partition
functional as
\begin{equation}
{\cal Z}=C(\beta)\int e^{-\int_0^{\beta}dx_4 \int d^3 x
L_{eff}} {\cal D}_i A_{\mu} {\cal D}\bar\psi {\cal D}\psi \delta (G) Det {\cal P}.
\end{equation}

Here $A_{\mu}$ is the electromagnetic field, $\bar\psi, \psi$ are fermion
fields, $C(\beta)$ is a normalization constant.
The gauge condition is $G$ and ${\cal P}$ is
the (trivial) Fadeev-Popov determinant
 \cite{Bernard},
\cite{Kapusta},\cite{Kalashnikov}.  We have also

\begin{equation}
L_{eff}=  L_{(\partial_{\nu} \to \partial_{\nu}-\mu\delta_{\nu 4})}
\end{equation}
We conclude that the fourth derivative of
fermions is shifted in $\mu$: the chemical potential
enter into the density matrix through the vector $c^{(1)}_{\nu} =\mu u_{\nu}$, and
the temperature through $c^{(2)}_{\nu}=T u_{\nu}$. Then, in the rest system, the field operators depend on the coordinate
"vectors"
$x_\nu=({\bf x},x_4)$, multiplied by the "momentum" vectors $P_{M\nu}=({\bf p},p_4)$, where $p_4$ are the Matsubara frequencies,
which are
 $2n \pi T$ for bosons and $(2n+1)\pi T$ for
fermions, where $n=0,\pm 1,\pm 2..$.  Thus, after taking the quantum statistical average
through the functional integration, indicated by the symbol $<<..>>$, we obtain
the thermodynamical potential $\Omega =-\beta ^{-1}\ln <<e^{-\int_0^\beta dx_4 \int d^3xL_{eff}(x_4, {\bf x)}}>>$.
We observe that the
statistical average leads to $<< \int dx_4\int d^3 x L_{eff}>> \to \Omega
$.

The energy-momentum tensor of matter
 plus vacuum will be obtained as a diagonal tensor
(no
shearing stresses occur in our approximation) whose spatial part
contains the pressures and the time component is minus the
internal energy density $-U$.
 The total energy-momentum tensor is obtained after quantum averaging as
\begin{equation}
{\cal T}^{\mu}_{ \nu }=<< T^{\mu}_{\nu }>>
\end{equation}
 where
\begin{equation}
T^{\mu}_{\nu}=
\frac{\partial  L} {\partial a_{i ,\mu }}a_{i ,\nu}
-\delta ^{\mu}_{\nu } L \label{ten}
\end{equation}

\noindent according to \cite {Landau}, where the index $i$ denotes the
fields (either fermion or vector components). In abscence of an external
field and/or a vector breaking the spatial symmetry, the only nonvanishing
contributions
from the terms $(\partial  L/\partial a_{i ,\mu })a_{i ,\nu}$ in (\ref{ten}) are those for which $\mu=\nu=4$. By considering the
fermion field contribution as example, we have, by integrating over $x_4,
{\bf x}$ the first term in the right of (\ref{ten})
\begin{equation}
\int dx_4 \int d^3 x\frac{\partial  L (\bar \psi (x)\psi(x),A_{\mu}(x))}
{\partial \psi_{,4}}\psi_{,4}=
\sum_n \int d^3p \bar
\psi(p_4,{\bf p})(i p_4\mu)\psi (p_4,{\bf p}) \label{ten1}
\end{equation}
the last expression leads to
\begin{equation}
\sum_n \int d^{3}p
(i p_4-\mu)\bar \psi(p_4,{\bf p}) \psi (p_4,{\bf p})=
\sum_n \int d^{3}p
[i p_4\frac{\partial L(p_4,{\bf p}))}{\partial ip_4}-\mu
\frac{\partial L(p_4,{\bf p})}{\partial \mu}]=\sum_n \int d^{3}p
[T\frac{\partial L(p_4,{\bf p})}{\partial T}-\mu \frac{\partial
L(p_4,{\bf p})}{\partial \mu}] \label{ten2}
\end{equation}
Performing the functional average of
(\ref{ten}) one gets $<<\int dx_4 \int d^3 x (\partial  L /
\partial \psi_{,4})\psi_{,4}>>
=T\partial \Omega/\partial T+\mu \partial \Omega/\partial \mu$. Then one
gets
the thermodynamical expression
\begin{equation}
{\cal T}^{i}_{j}=
-\Omega
\delta^i_j;\hspace{1cm}{\cal
T}^{4}_{4}=-(TS+\mu N+\Omega)=-U \label{tg}
\end{equation}
\cite{Landau1}, where ${\cal T}_{ij}=p \delta_{ij}$ are the
(isotropic) pressures,
$S=-\partial
\Omega/\partial T$ is the entropy density, $N=-\partial
\Omega/\partial \mu$ is the density of particles and $U$ the
internal energy density.

For black body radiation in equilibrium
\cite{Landau1} it is  ${\cal T}_{bj}^i=-\Omega_b \delta^{i}_{j}$,
where $i,j=1,2,3$, and ${\cal T}_{b4}^4=-U_b=3\Omega_b=-\pi^2
T^4/15\hbar^3 c^3$.

In the case when the Lagrangian depends  on a nonvanishing field
derivative, for instance, $a_{\mu,\lambda} \neq 0$,
then the first term in the right hand side of (\ref{ten}) is
nonzero and the pressures are given by ${\cal T}^{i}_{j}=(\partial
\Omega/\partial a_{i,\lambda})a_{j,\lambda}-\Omega \delta_{ij}$.
This happens when
there is an external field $a_{\mu}=A_{\mu}=B[-x_2, x_1,0,0]/2$
describing a
constant magnetic field (taken along the $3$-rd
axis),  which
generates non-vanishing
spatial tensor terms
through
the electromagnetic field tensor $<A^{\nu}_{,\mu}-A^{\mu}{,\nu}>={\cal
F^{\mu}_{\nu}}$.  This leads to pressure terms of form
${\cal T}^i_j=-\Omega \delta^i_j-{\cal F}^{i}_{k}(\partial
\Omega/\partial {\cal F}_{k}^{j})$, or
\begin{equation}
{\cal T}^{1}_{1}={\cal T}^{2}_{2}={\cal T}_{\perp}
=-\Omega-B{\cal M}
\end{equation}
where ${\cal M}=-\partial {\Omega}/\partial B$, is the
magnetization and $i=1,2$, $j=2,1$
For matter in an external magnetic
field, an anisotropy in the pressures
occurs \cite{Chaichian}, \cite{Shabad}. The anisotropy is due  to the
arising of a negative
transverse pressure,  generated by an axial
"force": the quantum analog of the Lorentz
force, arising when the magnetic field acts on charged particles having
non-zero spin
\cite{Aurora}, and leading to a magnetization parallel to
$\textbf{B}$.

\section{The vacuum limit. Casimir pressures}
Quantum statistics at temperature $T$ and
chemical potential $\mu$ leads to quantum field theory in vacuum
 if the limit limit $T \to 0$, $\mu \to
0$ is taken (see e.g. Fradkin \cite{Fradkin}; the functional
average $<<..>>$ becomes the quantum field average). This is
because the contribution of observable particles, given by the
statistical term $\Omega_{s}( T, \mu)$ in the expression for the
total thermodynamic potential $\Omega=\Omega_{s}+\Omega_{0}$,
vanishes in that limit. The remaining term, which is the
contribution of virtual particles, leads to the zero point  energy
of vacuum $\Omega_{0}$. Thus, we can find the total
energy-momentum tensor, and take at the end the limit $T \to 0$,
$\mu \to 0$ the quantum field average. If in (\ref{tg}) the
quantities $<a_{i,\mu}>=0$, then
\begin{equation}
{\cal T}^{i}_j=-\Omega \delta^i_j \label{isovac}
\end{equation}
is the isotropic
pressure and we conclude that {\it for the isotropic vacuum, if
the energy density $\Omega>0$, the pressures would be negative}
(and on the opposite, if $\Omega<0$, ${\cal T}_{ij}>0$). However,
this is not the case if there is a breaking of the spatial
symmetry (leading to momentum non-conservation; momentum
conservation is restored if we include the sources of the symmetry
breaking). These are just the case under study, and they are
especially interesting since they provide examples in which
divergences can be subtracted, leading to finite negative energy
terms, as seen before. The energy-momentum tensor for the Casimir
effect in vacuum may be obtained, however, directly by following a
complete parallelism with the temperature case, since the
four-momentum vector breaking the spatial symmetry has a discrete
component $p_3={\cal P}n$, $n=0,\pm 1,\pm 2..$. By choosing the
Lorentz gauge $\partial_{\mu}A_{\mu}=0$, one can write
\begin{equation}
{\cal T}^{3}_{3}=
<<\int_0^{d/\pi \hbar}dx_3\int d^{3}x'[ A_{\mu,3}\partial L/\partial
A_{\mu,3}-L]>>
\end{equation}
Where $d^{3}x'=dx_1 dx_2 dx_0$. After quantum averaging (only two degrees
of freedom are left;
$\Omega_{0C}$ is obtained in the Appendix, and coincides with
\cite{Casimir}), one has the anisotropic
pressures,
\begin{equation}
{\cal T}^{3}_{3}=P_{C3}={\cal P}\frac{\partial
\Omega_{0C}}{\partial {\cal P}}-\Omega_{0C} =
3\Omega_{0C}=-\frac{\pi^2 \hbar c}{240 d^4}<0 \label{Milon2}
\end{equation}
\noindent which is the usual Casimir negative pressure and
\begin{equation}
{\cal T}_{\perp}^C=P_{C\perp}= -\Omega_{0C} =\frac{\pi^2 \hbar
c}{720 d^4}>0 \label{Milon3}
\end{equation}
which is a  positive pressure acting parallel to the plates in the
region inside them. This is a second Casimir force. (This
\textit{is not} the so-called lateral Casimir force reported in
\cite{Mostepanenko}). The combined action of both forces suggests
a flow of QED vacuum out of the cavity inside the plates, as a
fluid which is compressed by the attractive force exerted between
them.

 One must remark that
the usual Casimir pressure corresponds to minus the energy of the
blackbody radiation at $T=T_{Cas}$, e.g., $P_{C3}={\cal T}^{C3}_{3}
\to {\cal T}^4_{b4}=-U_b$  and the Casimir energy corresponds to
minus the blackbody pressure ${\cal T}^{C4}_{4}= -\Omega_{0C}\to
-\Omega_b$, that is, both tensors are similar under exchange of
their ${\cal T}^3_3, {\cal T}^4_4$ components.

\section{The magnetized vacuum}
In the magnetic field case, according to
(\cite{Shabad}, \cite{Chaichian}), the diagonal
components
 lead to a positive pressure ${\cal T}_{3}^{0e3}=P_{03} =
-\Omega_{0e}$ along the magnetic field $B$, and to ${\cal
T}_{\perp}^{0e}=P_{0 \perp}= -\Omega_{0e}-B{\cal M}_{0e}$ in the
direction perpendicular to the field. Here ${\cal
M}_{0e}=-\partial\Omega_{0e}/\partial B$ is the vacuum
magnetization, which is obtained from (\ref{EH}) as
\begin{equation}\label{magnetization}
{\cal M}_{0e}=-\frac{2\Omega_{0e}}{B}- \frac{\alpha
B_c}{8\pi^2}\int_0^{\infty}e^{-B_c x/B}\left[\frac{coth x}{x}
-\frac{1}{x^2}-\frac{1}{3}\right]d x.
\end{equation}
\noindent One can check easily that (\ref{magnetization}) is a
positive quantity. Moreover, it has a non- linear dependence on
the field $B$. Then it may be stated that the quantum vacuum has
ferromagnetic properties (and ${\partial \cal M}_{0e}/\partial B
>0 $), although in our present one-loop approximation we do not
consider the spin-spin interaction between virtual particles.

 Concerning the transverse
pressure $P_{0e \perp}=-\Omega_{0e} -B {\cal M}_{0e} $, we get
\begin{equation}\label{pressure}
{\cal T}_{\perp}^{0e}=P_{0e \perp}=\Omega_{0e} + \frac{\alpha B_c
B}{8\pi^2}\int_0^{\infty}e^{-B_c x/B}\left[\frac{coth x}{x}
-\frac{1}{x^2}-\frac{1}{3}\right]d x.
\end{equation}
Both terms in (\ref{pressure})are negative, thus, $P_{0e
\perp}<0$, whereas along the field, the pressure
$P_{03}=-\Omega_{0e}$ is positive. This leads to magnetostrictive
effects for small as well for high fields. This could be tested by
placing parallel (non necessarily metallic) plates parallel to
$B$. Such plates would be compressed in the direction
perpendicular to $B$. Thus, QED vacuum in a magnetic field $B$ is
compressed perpendicular to it, and as the pressure is positive
along $B$, it is stretched in that direction. Obviously, for
non-metallic plates perpendicular to $B$, the pressures are
positive perpendicular to the plates and negative along them. This
is reasonable to expect: the virtual electrons and positrons are
constrained to bound states in the external field, but flows
freely in both directions along the field. That motion of virtual
particles can be interpreted as similar to the real electrons and
positrons, describing "orbits" having a characteristic radius of
order $\lambda =\sqrt{\hbar c/eB}$ in the plane orthogonal to $B$,
but the system is degenerate with regard to the position of the
center of the orbit. It must be stressed that the term $B{\cal
M}_{0e}$ subtracted by $-\Omega_{0e}$ in $P_{0e \perp}$ is the
statistical pressure due to the quantum version of the Lorentz
force acting on particles (in the present case virtual) bearing a
magnetic moment, which leads to ${\cal M}_{0e}>0$ \cite{Aurora}.
In the low energy limit $eB<<m^{2}$ we have $P_{0e \perp} \approx
3{\Omega_{0e}}<0$. It can be written

\begin{equation}\label{pv}
  P_{0e \perp}\approx
-\frac{\pi^{2} \hbar c}{1920 b^{4}},
\end{equation}
 For small $B$ fields
of order $10-10^3$G, $P_{0e \perp}$ is negligible as compared with
the usual Casimir pressure. But for larger fields, e.g. for $B\sim
10^{5}$ G it becomes larger; one may obtain then pressures up to
$P_{0e \perp} \sim10^{-9} dyn$ $ cm^{-2}$ . For a distance between
plates $d=0.1cm$, it gives $P_{0C} \sim 10^{-14}dyn$ $cm^{-2}$,
(see below) i.e., five orders of magnitude smaller than $P_{0e
\perp}$.  Our results show that quantum vacuum in a constant
magnetic field  may exert pressures, either positive or negative,
which means \textit{a transfer of momentum from vacuum to real
particles or macroscopic bodies} (as well as in Casimir effect,
see below). A similar idea is approached from classical grounds in
\cite{Feigel}.

\section{Discussion}
It is easy to check that for the magnetic field case, the
expression for the energy-momentum tensor of vacuum is
Lorentz-invariant with regard to inertial frames moving parallel
to $B$. It is also easy to check that the Casimir energy-momentum
tensor defined by (\ref{Milon}), (\ref{Milon2}) and (\ref{Milon3})
remains invariant with regard to Lorentz transformations to
inertial frames parallel to the plates.
The trace ${\cal T}^{0e\mu}_{\mu}=-4\Omega_{0e}-2B{\cal M}_{0e}$
whereas ${\cal T}^{C\mu}_{\mu}=0$. However, it is of especial
interest to consider the trace of the tensor ${\cal T}_{\mu
\mu}$ in both cases. In the classical isotropic case it is ${\cal
T}_{\mu \mu}=\rho +3p$. In the Casimir case one has
${\cal T}_{\mu \mu}^{C }=2\Omega_C <0$ and in presence of the
magnetic field it is ${\cal T}_{\mu \mu}^{0e}=-2\Omega_{0e}-2{\cal
M}B<0$. It is easy to check that in the magnetic field and Casimir cases
both the average pressure ${\cal T}_{ii}/3$
 and the energy density, are negative, since it  is
$\langle p_{0e}
\rangle=-\Omega_{0e}
+ 2B{\cal M}/3<0$, but $\langle p_{0e}
\rangle/\Omega_{0e}>1$, and for the Casimir case we obtain similarly
$\langle p_{0C}
\rangle=\Omega_{0C}/3<0$, (and $\langle p_{0C}
\rangle/\Omega_{0C}=1/3$ as in blackbody radiation).

Quantum vacuum energy has been suggested as a possible candidate to
dark energy, leading to a repulsive gravity, equivalent
to a cosmological constant \cite{Turner},\cite{Bucher}.
The condition $3p + \rho<0$ is expected to be fulfilled in Einstein equations assuming the energy density
$\rho>0$, and in consequence the average pressure $p<0$, which means $w=(p/\rho)<-1/3$. (This also might be
understood as a consequence of (\ref{isovac}) if $\Omega>0$).

In considering the effect of magnetized vacuum, we shall assume
the magnetic lines of force in intergalactic space as describing
curves  in all directions (there is no preferred direction for the
magnetic field $B$) , or either, we may assume that there are
magnetic domains randomly distributed in space, assuming in each
of them the magnetic field as constant, so as to give an isotropic
spatial average of the energy-momentum tensor. Concerning the
energy term, it must contain the contribution of the average density of the matter
creating the magnetic field.

We will refer especially to the extreme case of superdense matter
where it is magnetized so that the pressure transverse to $B$ is
$P_{\perp}= -\Omega -B{\cal M}$. For $-\Omega \leq B{\cal M}$, the
transverse pressure vanishes or becomes negative, leading to
unstable conditions: the gravitational pressure exerted by the
body cannot be balanced by matter pressure, the outcome being an
anisotropic collapse \cite{Chaichian}. For the energy density
under these conditions one can write $U \leq N\mu -B{\cal M}$.  We
observe that magnetic fields decrease the energy density, although
we expect that $U \geq 0$ in any case, i.e. under stable
conditions the (negative) magnetic energy never exceeds in modulus
the rest energy. (This happens  for a gas of charged vector
bosons, whose ground state has a decreasing but non-vanishing
effective mass $\sqrt{M^2 c^4-eBc\hbar}>0$ \cite{Elizabeth} (for
increasing $B$)). Under that assumption, the total energy density
would be $U \leq N\mu -B{\cal M}+\Omega_{0e}>0$. Thus, we expect
that even if the average magnetic vacuum pressure taken in large
regions of space have a negative sign, the average energy density
of magnetic field+sources should be positive. The situation  would
be similar to the case discussed after (\ref{isovac}). By
comparing (\ref{h}) with the estimated density of visible matter
(around $10^{-10}$erg/cm$^3$) \cite{Turner3}, it would mean a
field of $10^{5}$G. For present cosmology this is too large, since
the  estimates for the intergalactic magnetic fields are in the
range $10^{-6}-10^{-9}$ G \cite{Turner2}.

However, the mechanism is interesting since if we assume the existence of some regions
of space having a distribution of strongly magnetized
matter, it might be possible to have vacuum average negative pressures leading inside these regions
to $\rho + 3\langle p\rangle <0$, with $\rho>0$. The mechanism might be interesting also in connection to
alternative inflationary models.

The main teaching of the results discussed in this letter is that QED vacuum
under the action of external fields having axial symmetry,  after spatial averaging,
provide the way of extracting finite vacuum negative pressure
densities and negative average energies.

\section{Acknowledgements}
The authors thank H. Mosquera Cuesta and A.E. Shabad for
discussions. H.P.R. is especially indebted to G. Altarelli, M. Chaichian, C.
Montonen,  D. Schwartz and R. Sussmann for
enlightment comments.
\section{Appendix}
As pointed out, there is a
correspondence between the expressions for the energy momentum
tensor in Casimir effect and in the black body problem at
temperature $T=T_{Cas}$.  The Casimir energy (\ref{Milon}) can be
shown to be equivalent to the thermodynamic potential of black
body radiation at temperature $T$. We start from the generalized
propagator ${\cal
D}(\epsilon)=[p_4^2+p_1^2+p_2^2+p_3^2+\epsilon]^{-1}$, where
$\epsilon$ is equivalent to a squared mass term to be taken
 as zero at the end. For the photon propagator we demand ${\cal
D}_0={\cal
D}(\epsilon=0)$. We start from the quantity
\begin{equation}
\Omega_\epsilon=2(2\pi\hbar)^{-3}T\sum_s \int d^3 p \ln{\cal
D}(\epsilon). \label{A1}
\end{equation}
\noindent By taking the derivative with regard to $\epsilon$ one
obtain an expression which can be  summed over $s$ \cite{Fradkin}.
Then, after integrating the result over  $\epsilon$, and by taking
$\epsilon=0$ afterwards, leads to the expression
\begin{equation}
\Omega=\frac{2T}{(2\pi)^3\hbar^3 c^3}\int d^3 p[\ln
(1-e^{\omega/T})+\omega/T ] \label{A2}
\end{equation}
\noindent where $\omega=\sqrt{p_1^2+p_2^2+p_4^2}$. The logarithmic
term can be transformed by integrating by parts, leading to the
usual expression for the black body thermodynamical potential,
which is $\Omega(T)= -\pi^2 T^4/45\hbar^3 c^3 $ plus a divergent
term, independent from $T$. But if
$\partial\Omega_\epsilon/\partial \epsilon$ is first integrated
over $p_3$, and the resulting expression is integrated over
$\epsilon$, the final result is exactly the same as the Casimir
energy (\ref{Milon}) by replacing $T$ by $\hbar c/2d$ in
$\Omega(T)$.

\end{document}